\documentstyle[eqsecnum,pre,preprint,aps]{revtex}
\tightenlines

\newcommand{\be}{\begin{equation}}

\newcommand{\ee}{\end{equation}}
\newcommand{\bs}{\begin{mathletters}} 
\newcommand{\es}{\end{mathletters}} 

\newcommand{\baa}{\begin{eqnarray}}
\newcommand{\eaa}{\end{eqnarray}}

\newcommand{\ba}{\bs\begin{eqnarray}}
\newcommand{\ea}{\end{eqnarray}\es}

\newcommand{\bt}[1]{\bs\label{#1}\begin{eqnarray}}
\newcommand{\et}{\end{eqnarray}\es}

\newcommand{\figlab}[2]{\begin{figure}\caption{#2}\label{#1}\end{figure}}

\newcommand{\eq}[1]{Eq.~(\ref{#1})}

\newcommand{\paper}[6]{#1 , #2 #3 {\bf #4} , #5 (19#6)}

\newcommand{\refsec}[1]{Sec.~\ref{#1}}

\newlength{\www}

\begin{document}
\draft
\preprint{ULDF-TH-1/7/99}

\title{Green Function Simulation of Hamiltonian Lattice Models\\
with Stochastic Reconfiguration}

\author{Matteo Beccaria}

\address{
Dipartimento di Fisica dell'Universit\`a di Lecce, I-73100, Italy,\\
Istituto Nazionale di Fisica Nucleare, Sezione di Lecce}

\maketitle

\begin{abstract}
We apply a recently proposed Green Function Monte Carlo to the study
of Hamiltonian lattice gauge theories. This class of algorithms
computes quantum vacuum expectation values by averaging over a set of
suitable weighted random walkers. By means of a procedure called 
Stochastic Reconfiguration the long standing problem of keeping fixed
the walker population without a priori knowledge on the ground state
is completely solved. In the
$U(1)_2$ model, which we choose as our theoretical laboratory, 
we evaluate the mean plaquette and the vacuum energy per plaquette.
We find good agreement with previous works using model dependent 
guiding functions for the random walkers. 
\end{abstract}


\section{Introduction}
\label{sec:intro}

Lattice regularization of quantum field theories is a powerful
technique to compute non perturbative physical quantities. Its
application to quantum chromodynamics has led to accurate predictions
of the spectrum of quark and gluon states as well as of other
phenomena like finite temperature phase transitions.

The historical development of this active field of research developed along
two main streams: the Lagrangian approach proposed by
K. Wilson in 1974~\cite{Wilson} and the Hamiltonian formulation
derived by J. Kogut in 1975~\cite{U1}. The two approaches
are equivalent in the continuum limit but, for the purpose of
analytical or numerical investigations, they offer quite different
advantages.

The Lagrangian approach exploits Feynman's old idea of computing 
quantum partition functions as sums over classical
histories. Vacuum expectation values are obtained by evaluating
suitable statistical averages over long time trajectories. 
The basic object in this approach is the classical trajectory
whose full temporal evolution must be retained. Space and time are 
treated in a symmetric way and a field configuration is defined over a
space time discrete lattice. The identification of a $d$ dimensional
Euclidean quantum theory with a $d+1$ dimensional statistical model is
full. We remark that 
the Lagrangian approach is the current technique for the study of QCD
mass spectra and the most precise studies~\cite{Morningstar2} suggest
the advantage of working on anisotropic lattices with different
treatment of the spatial and temporal discretisation. In particular,
the use of spatially coarse, temporally fine lattices has greatly
increased the efficiency of the numerical simulations~\cite{Morningstar1}.

These considerations lead to a renewed interest in the Hamiltonian
formulation where only space is discretised and time
remains continuous. This is a very natural approach in the study of
low energy physics. Given the Hamiltonian $H$ of a quantum many body
model, its ground state is projected out by the application of the
evolution operator $U(t) = \exp(-t H)$, with $t\to +\infty$, to any
state with the same quantum numbers of the vacuum. Any accurate
representation of the asymptotic behaviour
of $U(t)$ provides access to the ground state  and to the low lying
excitations. These representations perform clever repeated
applications of $H$, in analytical or stochastic way, to specific
quantum states that replace in this approach the role played by
classical trajectories in the Lagrangian one.

One of the first striking examples of this strategy can be found 
in~\cite{Lanczos} where the authors obtain good scaling
measures of the mass gap in several non trivial models by analytical 
Lanczos diagonalization of the Hamiltonian. Many other analytical
techniques have been developed based on similar ideas like, for instance,
resummation of $U(t)$ expansions or variational approximations to the
spectrum. An updated list can be found in~\cite{ReviewHamiltonian}.

On the numerical side, powerful Monte Carlo algorithms exist for the
solution of quantum many body problems~\cite{Linden}. 
In this paper we are mainly concerned with those belonging to the so
called class of Green Function Monte Carlo (GFMC). 

They may be regarded as the lattice version of the 
Feynman-Kac representation~\cite{Simon} for matrix elements of
$U(t)$ that are computed by averaging over ensembles of
suitable random walkers. The dynamics of the walkers is determined 
by the {\em kinetic} part of the Hamiltonian that is its off diagonal matrix
elements in the basis of walker states.
The {\em potential}, the diagonal matrix elements of $H$, enters in
the definition of a path-dependent weight which the walkers carry to
represent their relative importance.
In such an approach, the problem is that the weights exponentially
explode or vanish as the walkers diffuse and after a short time the
algorithm becomes completely unfeasible.

To solve this problem there are two standard classes of solutions: (a)
the introduction of a guiding function for the random walkers, (b)
a branching mechanism for the walkers population. Within the former
method (see~\cite{GRW,Barnes,Hamer} for applications to lattice gauge
theories), 
the measure in the path space is deformed and the
simulation generates guided random walks according to a guiding
function which is inspired by exactly known properties of the ground
state (typically, its weak and strong coupling approximations). The
disadvantage of the method is that it requires tuning of the
functional form of the guiding function. This can only be achieved by 
a somewhat accurate knowledge of the vacuum structure. In fact,
perfect guidance is equivalent to the full problem of determining the
vacuum wave function.

In the latter solution, branching is introduced among the walkers to
damp their weight variance. Walkers with low weights are deleted and
relevant walkers are replicated. The problem in this case is a rather
involved management of the variable size walker population.

In a recent paper~\cite{GFMCSR}, a simple 
procedure called Stochastic Reconfiguration, has
been successfully applied to overcome this problem and perform
simulations with a fixed number of walkers without introducing any
guiding function, which can however be exploited if available.

The aim of this paper is to investigate the application of this
new technique to lattice gauge theories. In particular, we apply the
method to $U(1)_2$ lattice gauge theory as a simple theoretical
laboratory where it is easy to present the main ideas and discuss the
systematical errors introduced by the algorithm as well as their
control.

The plan of the paper is the following. In \refsec{sec:review} we
review the GFMC with Stochastic Reconfiguration. In
\refsec{sec:application} we apply the algorithm to the $U(1)_2$
model. We discuss its actual implementation, the
optimization of its free parameters and the numerical simulations.
Finally, in \refsec{sec:conclusions} we summarize our results and 
discuss the perspectives in the study of realistic gauge models.

\section{Review of GFMC with Stochastic Reconfiguration}
\label{sec:review}

The description of GFMC with Stochastic Reconfiguration is equivalent
to present the Feynman-Kac formula on a lattice. For this reason 
we begin by discussing the simple example of quantum mechanics in flat
space.
Let us consider the
one dimensional Schr\"odinger hamiltonian for a unit mass point particle
\be
H = \frac 1 2 p^2 + V(q) = H_0 + V(q)
\ee
and the problem of computing its ground state wave function.
We define a discrete Markov chain as follows. Let $\epsilon$ be
the time step (not necessarily infinitesimal) associated to each
Markov jump and let the state of the chain be specified by the
position eigenvalue $q$. Let the transition function $p(q'\to q'')$ be
\be
p(q'\to q'') = K_0(q'', q', \epsilon)
\ee
where $K_0$ is the propagator of the free hamiltonian $H_0$
\be
K_0(q'', q', t) = \langle q'' | \exp - t H_0 | q'\rangle = \frac{1}{\sqrt{2\pi t}} \exp-\frac{(q''-q')^2}{2t}
\ee
An ensemble of walkers can be described at step $n$ by its probability
density $P_n(q)$. It evolves according to 
\be
P_{n+1}(q'') = \int_{\bf R} dq'\ P_n(q') K_0(q'', q', \epsilon)
\ee
We can identify $P_n(q)$ with the wave function of the abstract
state $|P_n\rangle$ satisfying
\be
P_n(q) = \langle q | P_n\rangle
\ee
and therefore, as is well known, we obtain 
\be
|P_n\rangle = e^{-n\epsilon H_0} |P_0\rangle
\ee
for any finite $\epsilon$.

To make a similar construction with $H_0$ replaced by $H$ we need an
extension of the formalism. We consider a Markov chain where the state
is extended from $q$ to the pair $(q, \omega)$ where $\omega$ is a real weight whose
dynamics we shall describe in a moment.
The transition kernel is assigned as follows
\be
p(q'\omega' \rightarrow q''\omega'') = K_0(q'', q', \epsilon)
\delta\left(\omega''-\omega' e^{-\epsilon V(q')}\right)
\ee
with the correct normalization 
\be
\int_{\bf R} dq''\int_0^\infty d\omega''\ p(q'\omega' \rightarrow q''\omega'') = 1
\ee
In other words, at each discrete step, $q$ diffuses making a random
step with variance $\langle
(\delta q)^2\rangle = \epsilon$ as before and the
weight $\omega$ is multiplied by the exponential factor $\exp(-\epsilon V(q))$.

The probability distribution $P_n(q, \omega)$ at the n-th iteration evolves according to 
the equation 
\be
P_{n+1}(q'', \omega'') = \int_{\bf R} dq'\int_0^\infty d\omega'\ P_n(q', \omega')\ p(q', \omega'\rightarrow q'', \omega'')
\ee
To recover a wave function evolving according to $\exp(-t H)$ we must average over the weights
and introduce the function 
\be
\psi_n(q) = \int_0^\infty d\omega \omega P(q, \omega)
\ee
that satisfies
\be
\label{evolution}
\psi_{n+1}(q'') = \int_{\bf R} dq' \psi_n(q') e^{-\epsilon V(q')} K_0(q'', q', \epsilon)
\ee
If we now write $\psi_n$ in terms of the associated basis independent abstract state
\be
\psi_n(q) = \langle q | \psi_n\rangle
\ee
we easily see that
\be
|\psi_{n+1}\rangle = e^{-\epsilon H_0} e^{-\epsilon V(q)} |\psi_n \rangle
\ee
and therefore, by Trotter-Suzuki formula, in the limit $\epsilon\to 0$
we obtain 
\be
\lim_{\epsilon\to 0} \psi_{t/\epsilon}(q) = \langle q | e^{-t H} |\psi_0\rangle 
\ee
that is the Schr\"odinger evolution is completely reproduced.

The meaning of the previous manipulations is that we can numerically compute the Schr\"odinger
semigroup by averaging over walkers which diffuse according to the kernel $K_0$ and which carry 
an additional weight $\omega$. The weight takes care of the potential and is the actual relative
weight of the walkers.
As discussed in the Introduction, in the actual implementation 
of this method one has to face the rapidly increasing variance
of the weights of an ensemble of walkers as the evolution time goes by.

Stochastic Reconfiguration is a solution to this problem based on the
observation that there are many $P(q, \omega)$ giving rise to the same
physical wave function $\psi(q)$. In particular
\be
\label{sr}
\int_0^\infty d\omega\ \omega P(q, \omega) = \int_0^\infty d\omega \
\omega \widetilde P(q, \omega)
\ee
where 
\be
\widetilde P(q, \omega) = \delta(\omega-1) \int_0^\infty d\omega'\ \omega' \
P(q, \omega')
\ee
and $P(q, \omega)$ and $\widetilde P(q, \omega)$ give rise to
the same wave function $\psi(q)$. The difference between them is
hidden in the weight statistics that is not observable. The advantage
of choosing the representative $\widetilde P(q, \omega)$ among all the $P(q, \omega)$
associated to a given $\psi(q)$ is that it has exactly zero weight variance
by construction since all the weights are fixed to unity. 

The actual implementation of \eq{sr} in a numerical algorithm is
straightforward. Let us consider an ensemble of $K$ walkers
(characterized by their position eigenvalue and weight)
\be
{\cal E} = \{(q_k, \omega_k)\}_{k=1, \dots, K}
\ee
In the $K\to \infty$ limit we can associate to the ensemble ${\cal E}$
a unique well defined distribution function $P_{\cal E}(q, \omega)$. We now build a new ensemble
$\widetilde{\cal E}$ with $K$ walkers and with the property that when
$K\to\infty$ we have 
\be
P_{\widetilde{\cal E}}(q, \omega) = \widetilde{P}_{\cal E}(q, \omega)
\ee
The new ensemble is
simply built by extracting walkers from ${\cal E}$ and assigning them 
a unit weight. The walkers with position $q_k$ are extracted with
probability proportional to $\omega_k$. 
In other words we extract the new $K$ walkers from the set of old
states (the values $\{q_k\}$) with probabilities
\be
p_k = \frac{\omega_k}{\sum_k \omega_k}
\ee
Repetitions may occur during this multiple extraction; they change the
relative frequency of the old low and high weight walkers. 
To see this, let us consider an ensemble of walkers $\{(q_i,
\omega_i)\}_{i=1,\dots,K}$ where for simplicity we assume that all $q_i$
are distinct. We want to compute the statistics of the fractions 
$\nu_i = n_i/K$ of walkers with state $q_i$ in the reconfigured
ensemble. Since each new walker is extracted independently, the
probability of building a new ensemble with $n_i$ walkers in the state $q_i$ is
given by the multinomial distribution
\be
p(n_1, \dots, n_K) = \frac{K!}{n_1! \cdots n_K!}\ p_1^{n_1}\cdots p_K^{n_K}
\ee
The average number of walkers with state $q_i$ is 
\be
\langle n_i\rangle  = \sum_{n_1 + \cdots + n_K = K} \frac{K!}{n_1! \cdots
n_K!}\ p_1^{n_1}\cdots p_K^{n_K}\ n_i = K p_i
\ee
and the mean product $n_i n_j$ with $i\neq j$ 
\be
\langle n_i n_j\rangle = K(K-1) p_i p_j
\ee
We conclude that 
\be
\langle \nu_i \rangle  = p_i,\quad
\langle \nu_i\nu_j \rangle -\langle \nu_i\rangle \langle \nu_j\rangle = -\frac{1}{K}
p_i p_j
\ee
Hence, in the reconfigured ensemble the state $q_i$ appears with a
frequency which is precisely $p_i \sim \omega_i$. Finite size correlations between
the new walkers are present vanishing with $K$ as $K^{-1}$. They
induce systematic errors in the measurements that must be eliminated
by extrapolation to the limit $K\to\infty$.

We conclude this Section with a comment about Importance Sampling. 
The aim of this paper is to show that it
is possible to perform simulations with a fixed size population of
walkers without any a priori knowledge of the vacuum state
$|0\rangle$. However, it must be emphasized that whenever a trial wave function for
$|0\rangle$ is available,  it can be easily included in the
algorithm by a unitary tranformation of $H$ as discussed in~\cite{IS}.
This is straightforward in the Hamiltonian formulation and therefore 
analytical approximations (e.g. variational calculations)
can be exploited to accelerate convergence.

\subsection{Generalization to other models}

From the above discussion, it should be clear that the extension of GFMC
with Stochastic Reconfiguration to a more general Hamiltonian is
possible only when some structural conditions are true. To be definite
we require the existance of a complete explicit basis $\{|s\rangle\}$ 
such that $H$ can be written
\be
H = T+V
\ee
where: (a) $V$ is real and has vanishing off diagonal matrix elements,
(b) $T$ is
the generator of a Markov process. 
Condition (b) is simply the statement that the evolution operator
$U(t) = \exp(-t\ T)$ exists for $t>0$ and the kernel
\be
K(s'', s', t) = \langle s'' | \exp(-t\ T) | s'\rangle
\ee
is positive and normalized 
\be
\sum_{s''} K(s'', s', t) = 1
\ee
In more physical terms, we require to be able to write the Hamiltonian 
as a potential term plus a {\em good} kinetic term like, on a generic
manifold, the Laplace-Beltrami operator which is associated to random
walks on the manifold. Interesting models which belong to this class
in the Hamiltonian formulation are the non linear $O(N)$ $\sigma$ model and the $SU(N)$ pure gauge
theories. 

\subsection{Observables Measurements}
\label{sec:Q}

In this Section we discuss how observables can be measured.
The ground state energy is the simplest observable to be
computed. To measure it, we take two arbitrary states $|\chi\rangle$ and $|\psi\rangle$
with non zero overlap with the ground state $|0\rangle$ and write
\be
E_0 = \lim_{t\to + \infty}
\frac{\langle \chi | H e^{-t H} |\psi\rangle}{\langle \chi | e^{-t H}
| \psi\rangle }
\ee
The state $|\psi\rangle$ describes the statistics of the initial state
of the chain. 
A convenient choice, although not always optimal, is to take for
$|\psi\rangle$ one of the walker states and to start accordingly the
Markov chain always from the same state. 
About $|\chi\rangle$, a good choice is to take the zero
momentum state which is annihilated by the kinetic (off-diagonal) part
of $H$. In other word, this is the generalization (on non flat
manifold) of the constant wave function. In this case we drop the
kinetic part of $H$ and obtain
\be
E_0 = \lim_{t\to + \infty}
\frac{\langle \chi | V e^{-t H} |\psi\rangle}{\langle \chi | e^{-t H}
| \psi\rangle }
\ee
where $V$ is the diagonal part of $H$. From the point of view of the
algorithm, the projection over $|\chi\rangle$ is nothing but 
the recipe of summing over all the walkers with
no additional final state weight.

These prescriptions may be refined in model dependent ways, but we
shall see that, 
at least in the model under consideration, they work properly.

Concerning other operators, let us analyze in some details the measure
of vacuum expectation values of operators $Q$ which are diagonal in the chosen
basis (the position eigenstates in the Schr\"odinger example).
They can be computed as the limit 
\be
\frac{\langle 0 | Q | 0\rangle}{\langle 0 | 0\rangle} =
\lim_{\tau\to +\infty} {\cal O}(\tau)
\ee
where 
\be
Q(\tau) = \lim_{t\to +\infty} 
\frac{\langle\chi | e^{-\tau H} Q e^{-t H} | \psi\rangle}
{\langle\chi | e^{-(\tau+t) H} | \psi\rangle} = \langle 0 | Q | 0
\rangle + {\cal O}(e^{-\tau(E_1-E_0)})
\ee
where $E_1$ is the energy of the first excited state $|1\rangle$ with
non vanishing matrix element $\langle 1 | Q| 0\rangle$.
In the above expression, the limit over $t$ is performed automatically
by the running of the Markov chain. It can be considered reached as soon
as equilibrium in the chain is achieved. The second limit requires
some care and to evaluate it we allow the walkers to diffuse 
for an additional time $\tau$ after the measurement of the observable. 
In the following Sections we shall check the rapid exponential convergence of 
$Q(\tau)$ with increasing $\tau$. Of course, from the $\tau$
dependence of $Q(\tau)$ the (finite volume) mass gap $E_1-E_0$ can be extracted.

\section{Application to the $U(1)_2$ Model}
\label{sec:application}

In this Section we make the previous discussion more detailed by
examining a specific example, the $U(1)_2$ lattice gauge model.
Its Hamiltonian is given by 
\be
H = \sum_{p=1}^L\left[ 
\left(\sum_{l_p=1}^3 -\frac{1}{2\beta}\frac{\partial^2}{\partial \theta_{l_p}^2}
\right) + \beta (1-\cos\phi_p)
\right]
\ee
where the link phases $\theta_{l_i}$ are defined in Fig.~1 and the
gauge invariant plaquette phase $\phi_p$ is 
\be
\phi_p = \theta_{1, p}+\theta_{2, p+1}-\theta_{3, p}-\theta_{2, p}
\ee
We assume periodic boundary conditions.
This lattice model has no continuum limit because its correlation
length in lattice units remains finite for all values of the coupling
$\beta$.
Nonetheless, it shares many features with the more realistic
models in higher dimensions and serves to illustrate the method in
an easy case.
We are interested in the numerical calculation of the ground state energy per plaquette $E_0/N$
and of the mean plaquette
\be
\langle U_p\rangle \equiv \langle 0 | \cos\phi_p |0\rangle
\ee
For these two observables, the first terms in the weak and strong coupling expansions are known~\cite{Barnes}
\be
\frac{E_0}{L} = \left\{ 
\begin{array}{l}
\beta-\frac 1 4 \beta^3 + \frac{89}{3840} \beta^7 + {\cal O}(\beta^{11}) \\ \\
0.9833-0.1209 \beta^{-1} + {\cal O}(\beta^{-2}) \qquad (N\ge 5)
\end{array}
\right.
\ee

\be
\langle U \rangle = \left\{ 
\begin{array}{l}
\frac 1 2 \beta^2-\frac{89}{960} \beta^6 + {\cal O}(\beta^{10}) \\ \\
1-0.4917 \beta^{-1} + {\cal O}(\beta^{-2}) 
\end{array}
\right.
\ee

\subsection{Optimization of the Algorithm}

Let us discuss in this Section the tuning of the algorithm. Its free
parameters are
\be
\epsilon,\quad K, \quad r, \quad \tau
\ee
where $\epsilon$ is the time step in the application of
$\exp(-\epsilon H)$, $K$ is the size of the walkers ensemble, $r$ is the
number of Markov chain steps between two reconfigurations and $\tau$
is the time which we discussed in \refsec{sec:Q}.
The best choice of $\tau$ of course depends on the chosen
observable. In principle, the same holds also for the first three
parameters, but for simplicity, we shall discuss their optimization
on general grounds independently on $Q$.

The parameter $\epsilon$ sets the scale of the elementary fluctuations
of the link phases $\theta$. Since $\langle(\delta\theta)^2\rangle =
\epsilon$ we require $\sqrt{\epsilon}\sim 2\pi/n_\theta$ with large
$n_\theta$ to obtain a good approximation of the continuum
diffusion. In our simulations we choose the conservative value
$n_\theta=50$ and therefore $\epsilon=0.015$.

The value of $K$ cannot be fixed. Instead, it must be varied and an
extrapolation to $K=\infty$ is needed. The following 
functional form for all $K$ dependent quantities
\be
\label{functional}
Q(K) =  Q(\infty) + \frac{c}{K^\alpha} + \cdots
\ee
turns out to be enough general to reproduce quite well the $K$
dependence. In our simulation we extract the three constants $Q(\infty)$, $c$
and $\alpha$ from $Q(K)$ at $K=10$, 100, 1000, 5000.

The parameter $r$ controls the frequency of the reconfiguration
process. A small $r$ is useless and expensive. Actually,
reconfiguration must be performed only when the walkers begin to show
a significantly large weight variance. In that case, reconfiguration
is effective. If the variance is small, then
reconfiguration reshuffles the ensemble without useful effects. On the
other hand, if $r$ is taken too large, reconfiguration with a finite
$K$ will destroy the information contained in the ensemble and the
systematic error at fixed $K$ will be large.
Our proposal for a choice of an optimal $r$ is to fix it by looking at the integrated
autocorrelation time of a relevant observable, say the vacuum
energy. We set $r=1$ and run the algorithm: the energy measurements
$\{E_i\}$ show an exponential decorrelation
\be
\overline{E_n E_m}-\overline{E}^2 \sim A e^{-|n-m|/\tau}
\ee
where as usual bars denote the average over the measurements.
We then set $r\equiv \tau$. This procedure has two advantages: (a) the
energy measurements taken after each reconfiguration are decorrelated,
(b) the autocorrelation $\tau$ is independent on $K$.
Statement (a) holds by construction. Property (b) follows from the
fact that $\tau$ is an intrinsic feature of the chaotic evolution of
each single walker and has no reason to be $K$ dependent as we have
checked explicitly. This is a useful property for the $K\to\infty$
extrapolation. 

In our simulations we consider the $U(1)_2$ model at $\beta =$ 
0.5, 1.0, 1.5, 2.0, 2.5 with $L=8$ to
reproduce the results of~\cite{Barnes}. In this case we find that 
the optimal $r(\beta)$ is roughly
\be
r(0.5) = 40,\ r(1.0) = 30,\ r(1.5) = 25,\ r(2.0) = 25,\ r(2.5) = 20
\ee

We now present our results for the ground state energy per plaquette
and mean plaquette. For each data point we performed $10^5$ Markov
chain steps.

\subsection{Measure of $E_0$}

We have computed the vacuum energy per plaquette at the four values
$K=10$, $100$, $1000$ and $5000$ on a $L=8$ spatial lattice. A fit with the functional form in
\eq{functional} determines the $\beta$ dependent exponent
$\alpha(\beta)$. In Fig.~(\ref{fig:E8}) we show the linear fit of the
numerical measures plotted as functions of $1/K^{\alpha(\beta)}$. The
results of the fit are collected in Tab.~(\ref{tavolaE8}) and compared
with the weak and strong coupling expansions. The results are quite
good, especially because the strong coupling series is only known up
to the next to leading term. The CPU time required for each point is
about one hour on a Pentium 200 processor. Moreover, we stress that to
obtain these results we only needed a very simple tuning of the
algorithm and no additional knowledge of the ground state
properties. As can be seen from the figure, the extrapolation is very
near the value obtained with the largest $K$ used.
In the present model, it is fairly easy to obtain analytical approximations
for the ground state energy per plaquette. The hamiltonian written in terms
of the gauge invariant phases $\phi_p$ is 
\be
H = \sum_{p=1}^N\left[\frac{1}{\beta}\left(-2\frac{\partial^2}{\partial\phi_p^2}+
\frac{\partial^2}{\partial\phi_p\partial\phi_{p+1}}\right)+\beta(1-\cos\phi_p)\right],
\ee
and the independent plaquette approximation gives 
\be
\frac{E_0^{(Mathieu)}}{L} = \beta+\frac{1}{2\beta} a_0(-\beta^2),
\ee
where $a_0(q)$ is the lowest characteristic value for the even solutions of the Mathieu equation
\be
y''(x) + (a_0(q)-2q\cos(2x))y = 0.
\ee
A more practical approximation can be obtained by a variational calculation based on 
the following simple independent plaquette Ansatz 
\be
\psi_0(\phi_1, \dots, \phi_N) = \exp\left(\lambda\sum_{p=1}^N \cos\phi_p\right).
\ee
In this case we obtain 
\be
\frac{E_0^{(var)}(\beta)}{L} = \beta + \frac 1 \beta \min_\lambda \left[(\lambda-\beta^2) 
\frac{I_1(2\lambda)}{I_0(2\lambda)}\right],
\ee
(where $I_n$ is the n-th modified Bessel function).
In Tab.~(\ref{tavolaE8}), columns {\em Mathieu} and {\em var} show the numerical 
values of these approximations.

As one can see, they appear to reproduce well the Monte Carlo data over the whole 
region of the coupling. This holds true especially for the estimate computed in terms of 
Mathieu functions. However, we stress again 
that the Monte Carlo extrapolated data have no systematic errors and have been obtained 
with the minimum a priori knowledge. As stated above, any analytical information on the 
ground state, like a variational ground state wave function, can be used for the acceleration of the Monte
Carlo simulation by including in the algorithm an importance sampling step.

\subsection{Measure of $\langle U_p\rangle$}

The results obtained for the average plaquette are quite similar to
the previous ones. Again, we show in Fig.~(\ref{fig:P8}) the linear
extrapolation after the determination of $\alpha(\beta)$. In
Tab.~(\ref{tavolaP8}) we see that the Monte Carlo data is well
matching the analytical series except in the region 
around $\beta=1.0$. However, the weak coupling series is not to be
trusted at $\beta=1.0$ since its plot shows a steep variation between
$\beta=1.0$ and $\beta=1.5$ signalling the need for an additional
term. The point at $\beta=1.5$ deviated by about 3\% with respect to
the strong coupling series. This can simply explained by assuming that
it is a too small value for strong coupling to apply and moreover the
next points at larger $\beta$ match better, below the percent level.
In Fig.~(\ref{fig:tau}) we show the dependence of $\langle
U(\tau)\rangle$ on the time $\tau$ to check the exponential
convergence to the vacuum expectation value.
Again, we stress that these numbers are obtained without a priori
information and over the whole coupling variation.

To summarize our results we collect in Fig.~(\ref{fig:summary}) a
graphical comparison of Monte Carlo data with the asymptotic $\beta\to 0$ and
$\beta\to \infty$ expansions.

\subsection{Computational Cost}

In this Section we address the problem of estimating the computational
cost of the algorithm. A realistic computation must follow two steps:
\begin{enumerate}
\item[a)] extrapolation $K\to\infty$ at fixed coupling $\beta$: 
\item[b)] continuum limit $\xi(\beta) \to\infty$  where $\xi$ is the
correlation length in lattice units (we do not discuss finite volume
effects and their control).
\end{enumerate}
In performing step (a) we need the computational cost as a function of
$K$ that is the CPU time required to achieve a give statistical error
$\epsilon_{stat}$ in the evaluation of an observable $Q$. With $N$
Markov chain steps we have 
\be
\epsilon_{stat} = \frac{\sigma(\beta, K)}{\sqrt{N/r(\beta)}}
\ee
where $\sigma_Q(\beta, K)$ is the standard deviation of $Q$
measurements and $N/r(\beta)$ is the number of independent
measurements. The dependence of $\sigma_Q$ on $K$ is simple. Due to a
self average effect we expect $\sigma_Q(\beta, K) =
\tilde\sigma_Q(\beta) K^{-1/2}$. Our explicit numerical simulations 
confirm this scaling law. The CPU time is roughly proportional to $NK$
since the reconfiguration process turns out to be only a small
fraction of the computational cost. Therefore
\be
\epsilon_{stat} \sim \frac{\tilde\sigma(\beta)}{\sqrt{T_{CPU}/r(\beta)}}
\ee
and $K$ cancels. The increase in CPU time associated to the management
of the ensemble is compensated by the larger statistics. In other
words, the cost of stochastic reconfiguration is just the cost of the
fit described by \eq{functional}: it is proportional to $p$ the number
of $k$ values used to extract $Q(\infty)$ from $Q(K)$.

The cost of step (b) is expected to be much more model dependent.
As the continuum limit is approached, the quantities which vary 
with $\beta$ are $\tilde\sigma(\beta)$, $r(\beta)$ and
$\alpha(\beta)$. There is no mechanism in this algorithm to prevent 
slowing due to increasing $r(\beta)$ when the correlation length
increases and the behaviour of the algorithm must be studied in the
case of realistic models with diverging correlation length. What we
observe in the considered model is a very mild $\beta$ dependence of the
exponent $\alpha(\beta)$ suggesting that 
\eq{functional} may hold without the need for a huge $K$. Moreover, an advantage of the
Hamiltonian formulation is that this slowing could be eventually 
reduced by importance sampling as discussed in the end of \refsec{sec:review};
in fact, preliminary investigations show that the use of a good trial wave function for 
the ground state can strongly reduce $\tilde\sigma_Q(\beta)$ 
and allow to perform simulations with reasonable errors using relatively small values
of $K$.

\section{Conclusions and Perspectives}
\label{sec:conclusions}

In this paper we applied a recently proposed many body Monte Carlo algorithm of
Green Function type to the study of lattice gauge models in
Hamiltonian form. This kind of algorithms compute the vacuum wave
function and many related quantities by averaging over a set of
suitable weighted random walkers. The method that we discussed solves
the annoying long standing problem of fixing the walkers number
without exploiting a priori informations on the ground state structure
and is thus quite general. The trade off is the introduction of a
systematic error, but we showed by explicit simulations in the
$U(1)_2$ model that it can be kept completely under control. The
algorithm can be applied to more interesting models like the non
linear $O(N)$ $\sigma$ model or $SU(N)$ pure gauge on which work is in
progress~\cite{Moro}. The perspectives of this works seem interesting
expecially in view of the recently proposed improved lattice
Hamiltonians~\cite{Improvement} for pure gauge models where we plan to investigate the
algorithm efficiency.

\acknowledgements
I thank Prof. G. Jona-Lasinio for many discussions on the Feynman-Kac
formula and its related quantum many body algorithms.
I also thank Prof. G. Curci for many useful discussions on the 
numerical simulation of lattice field theories.
Finally, financial support from INFN Iniziativa Specifica RM42 is acknowledged.

\figlab{fig:1}{$U(1)$ N-chain geometry. Definition of the link angles.}
\figlab{fig:E8}{Extrapolation $K\to \infty$ for the vacuum energy per plaquette.}
\figlab{fig:P8}{Extrapolation $K\to\infty$ for the mean plaquette.}
\figlab{fig:tau}{Large $\tau$ limit of $\langle U(\tau)\rangle$.}
\figlab{fig:summary}{Graphical comparison of Monte Carlo data with strong and
weak coupling series.}

\begin{table}
\caption{$E_0/L$ compared with weak and strong coupling approximations. The last column shows the relative deviation
between the Monte Carlo (MC) data and the weak or strong expansions.}
\label{tavolaE8}
\begin{center}
\begin{tabular}{ccccccc}
$\beta$ & weak & MC & Mathieu & var & strong & \% $\Delta$ \\
\tableline 
0.5 & 0.4689 & 0.4694(2) & 0.4690 & 0.4690 & --- & 0.1  \\
1.0 & 0.7732 & 0.7736(2) & 0.7724 & 0.7746 & --- & 0.05 \\
1.5 & --- &    0.8907(2) & 0.8909 & 0.9005 & 0.9027 & 1    \\
2.0 & --- &    0.9292(2) & 0.9299 & 0.9435 & 0.9229 & 0.7  \\
2.5 & --- &    0.9524(2) & 0.9466 & 0.9594 & 0.9349 & 2    
\end{tabular}
\end{center}
\end{table}

\begin{table}
\caption{$\langle U\rangle$ compared with weak and strong coupling approximations.}
\label{tavolaP8}
\begin{center}
\begin{tabular}{ccccc}
$\beta$ & weak & MC & strong & \% $\Delta$ \\
\tableline 
0.5 &  0.1236 & 0.1229(2) & ---    & 0.6  \\
1.0 &  0.4073 & 0.4262(2) & ---    & 5 \\
1.5 & ---    & 0.6513(2)  & 0.6722 & 3    \\
2.0 & ---    & 0.7590(2)  & 0.7542 & 0.6  \\
2.5 & ---    & 0.8074(2)  & 0.8033 & 0.5 
\end{tabular}
\end{center}
\end{table}

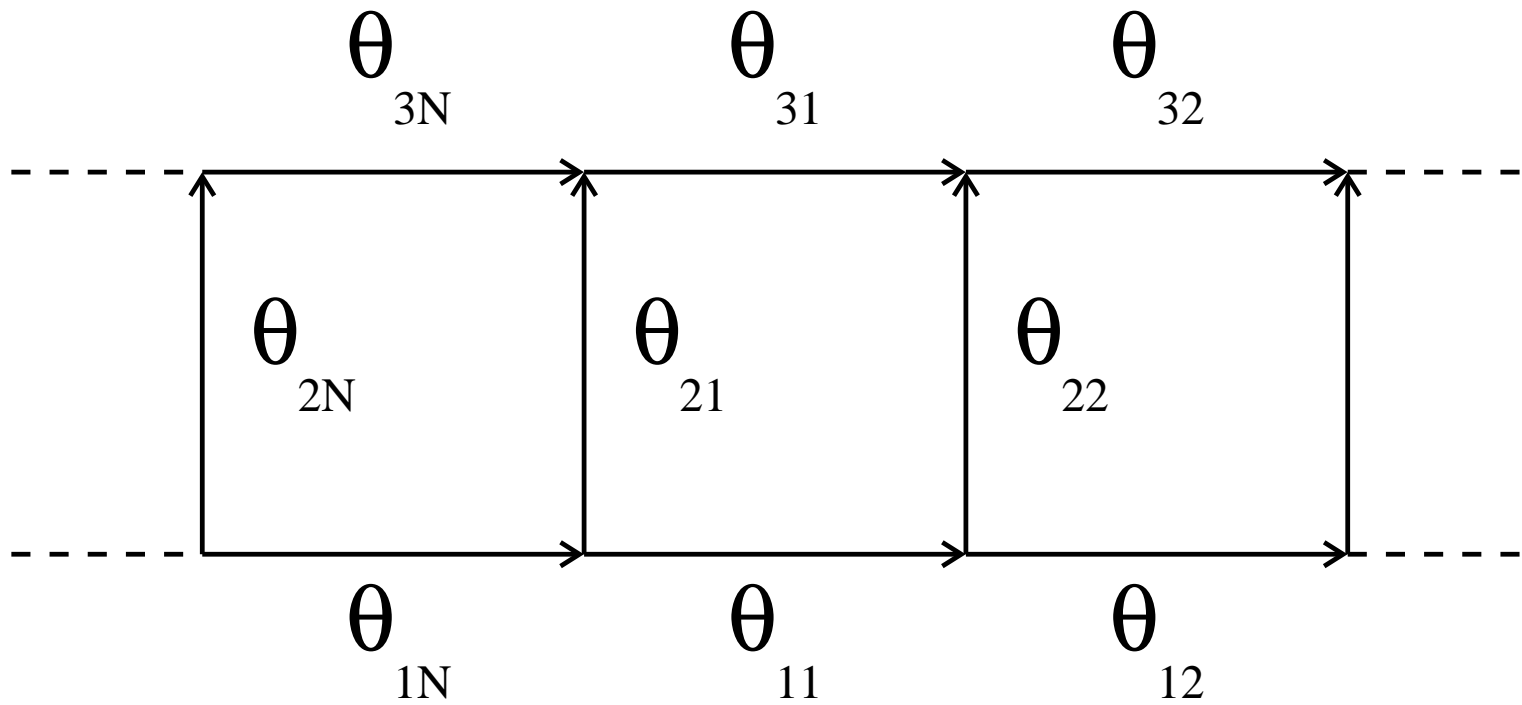
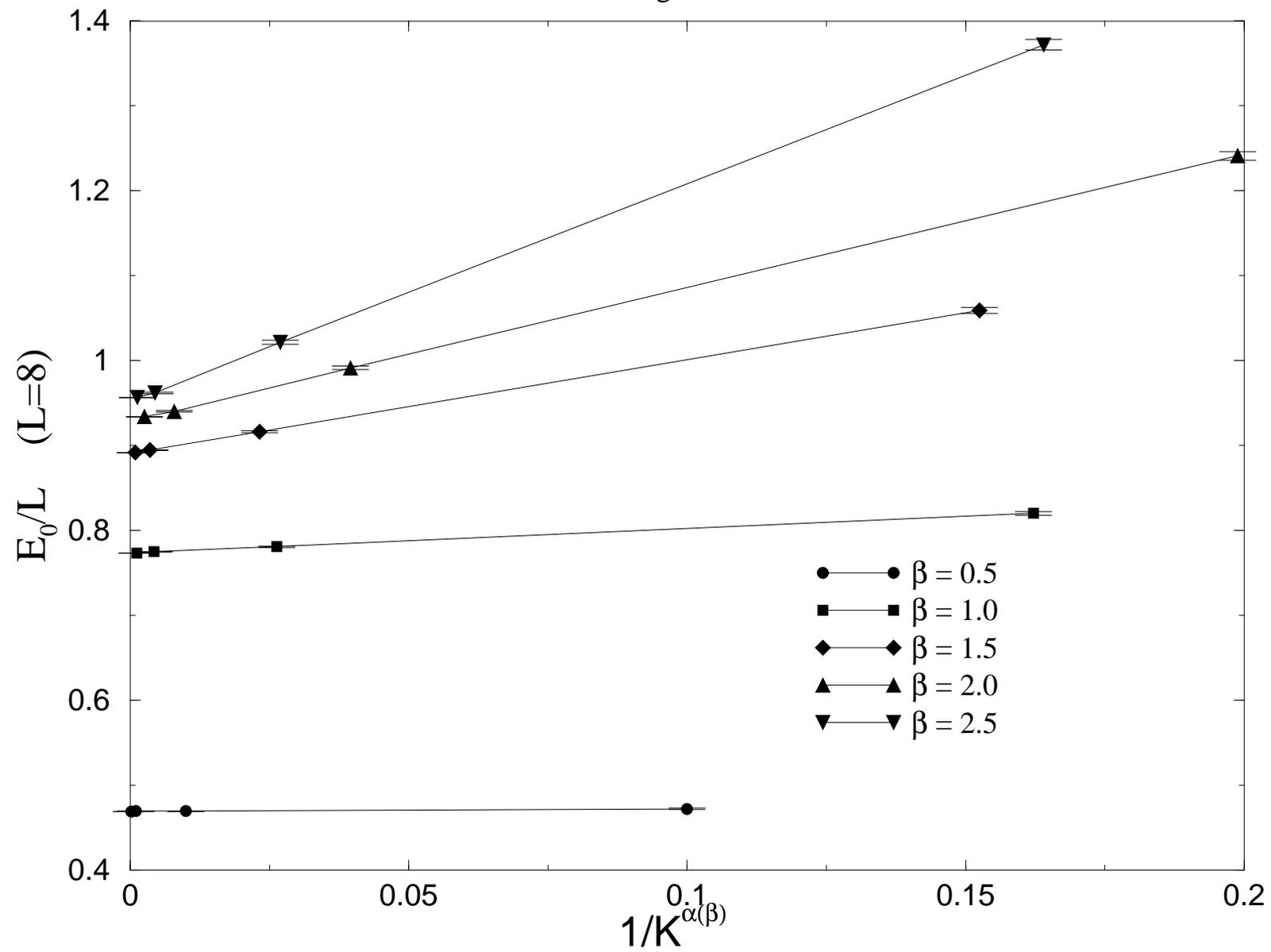
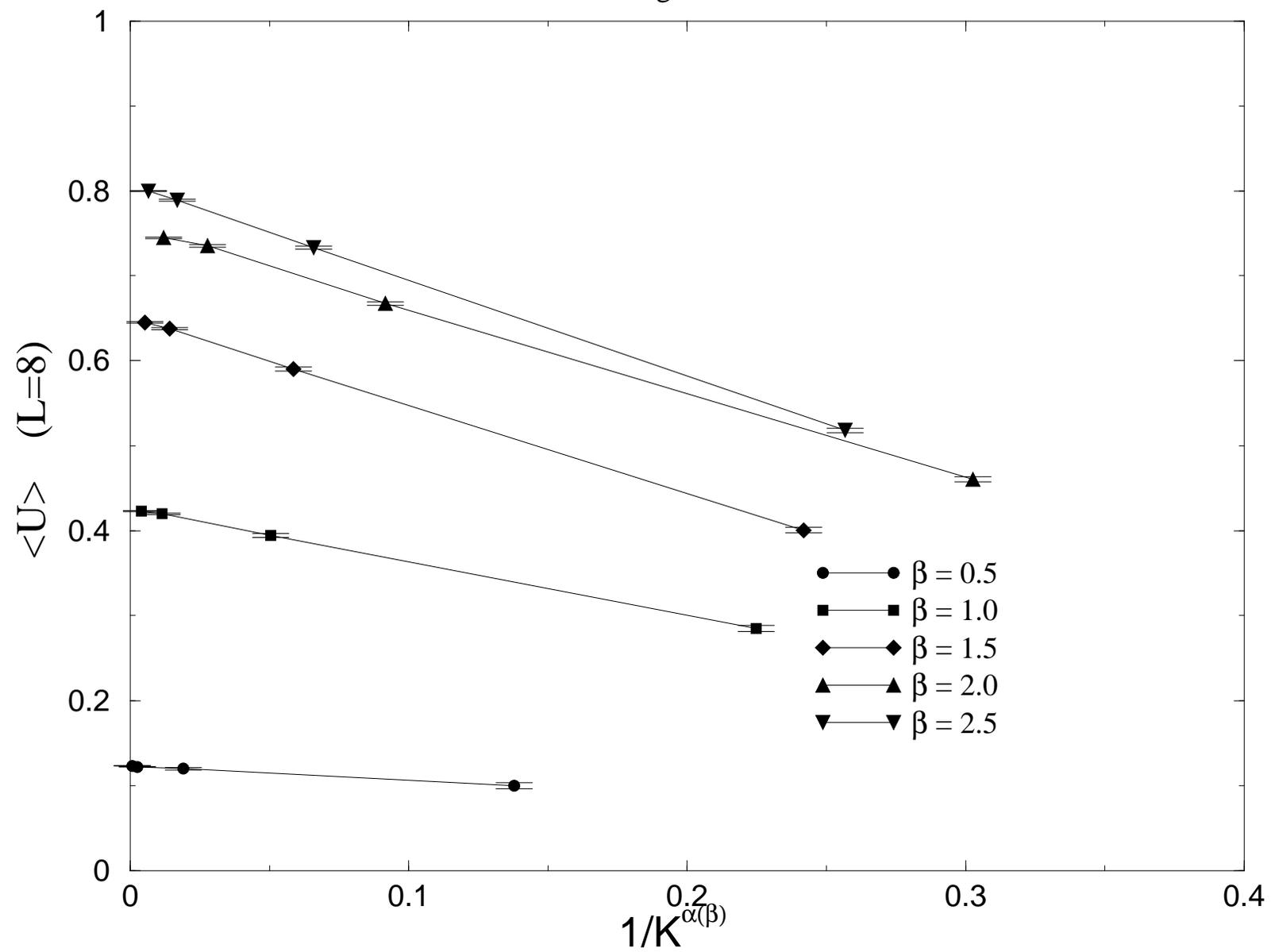
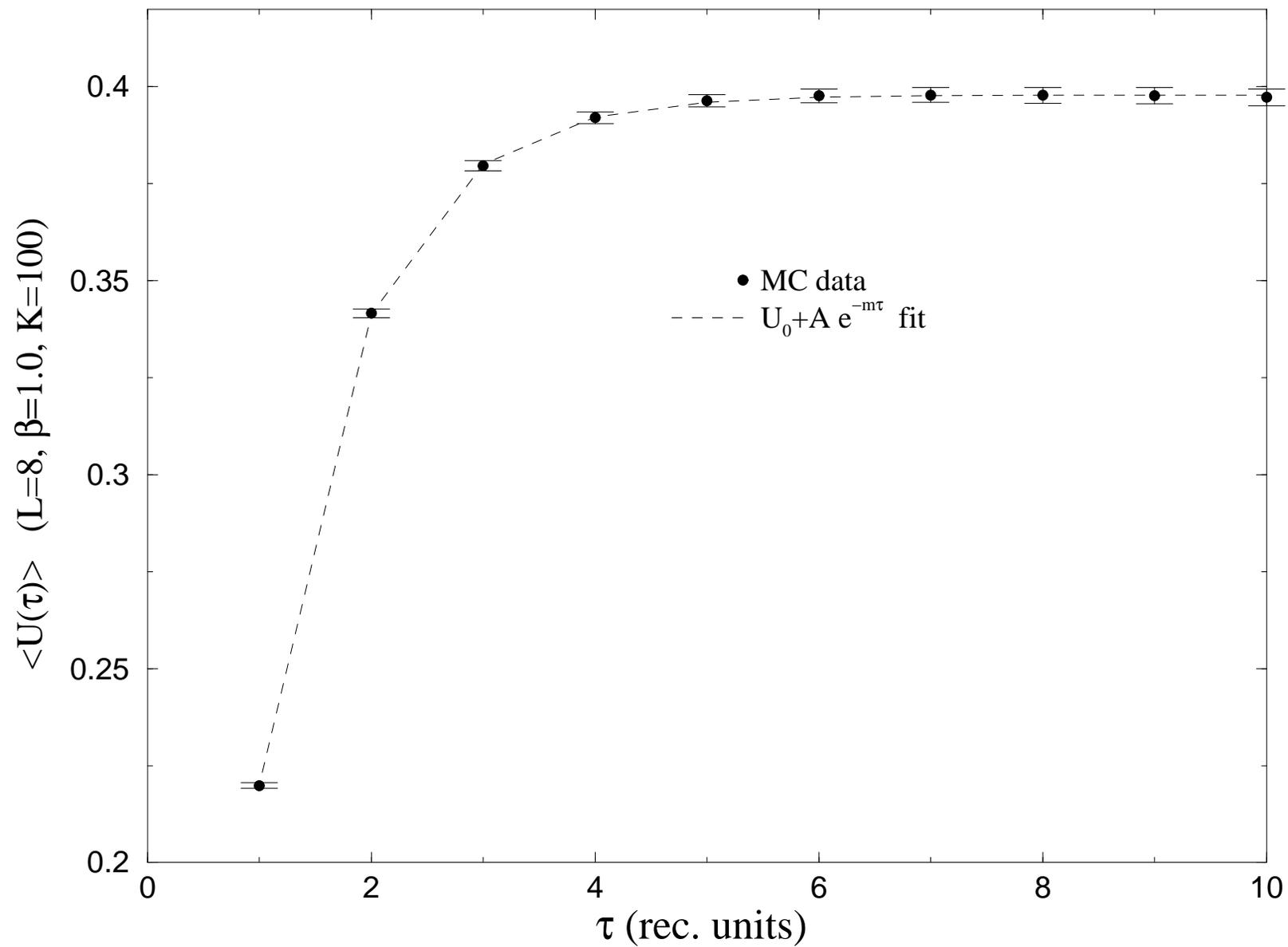
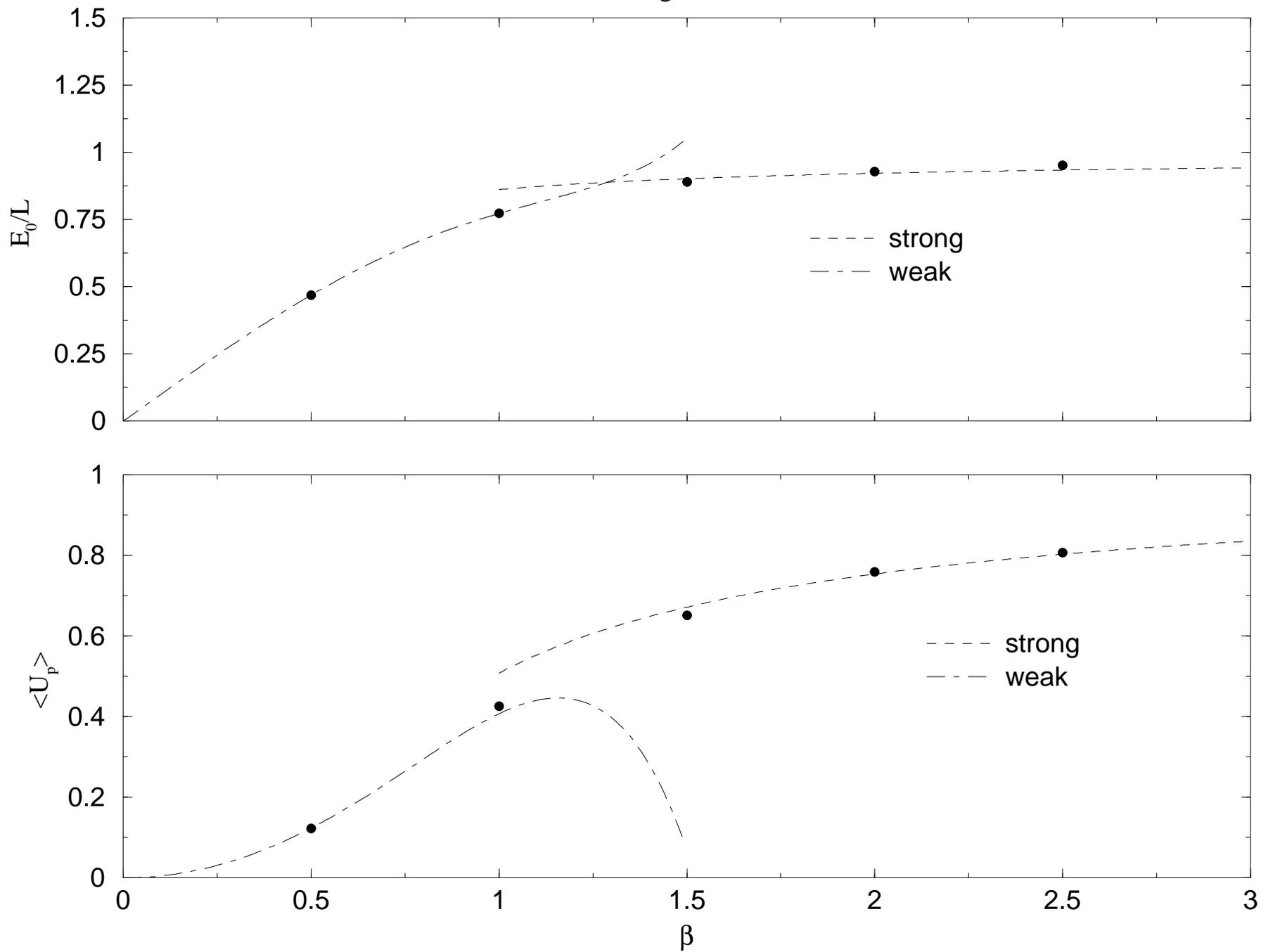
\end{document}